\documentclass[draftclsnofoot,11pt,onecolumn]{IEEEtran}
\usepackage{amssymb}
\usepackage[dvips]{graphicx}
\usepackage[cmex10]{amsmath}
\usepackage{amsthm}
\usepackage{latexsym,bm}
\usepackage{color}
\usepackage{subfigure}
\usepackage{multirow}
\usepackage{citesort}
 \def\bh{{\mathbf{h}}}

\def\bB{{\mathbf{B}}}

\theoremstyle{plain}

\theoremstyle{definition}
\theoremstyle{definition}

\begin{document}

\title{Performance Analysis of Protograph LDPC Codes for Nakagami-$m$ Fading Relay Channels}

\author{\IEEEauthorblockN{\fontsize{11pt}{\baselineskip}\selectfont {Yi Fang$^{1,2}$, Kai-Kit Wong$^{2}$, Lin Wang$^{1}$, and Kin-Fai Tong$^{2}$}}
\IEEEauthorblockA{\normalsize{
\\
$^1$Department of Communication Engineering, Xiamen University,
Xiamen, P. R. China\\
$^2$Department of Electronic and Electrical Engineering, University College London, London, UK\\
E-mail: wanglin@xmu.edu.cn
}}}

\maketitle

\begin{abstract}
In this paper, we investigate the error performance of the protograph (LDPC) codes over Nakagami-$m$ fading relay channels. We first calculate the decoding thresholds of the protograph codes over such channels with different fading depths (i.e., different values of $m$) by exploiting the modified protograph extrinsic information transfer (PEXIT) algorithm. Furthermore, based on the PEXIT analysis and using Gaussian approximation, we derive the bit-error-rate (BER) expressions for the error-free (EF) relaying protocol and decode-and-forward (DF) relaying protocol. We finally compare the threshold with the theoretical BER and the simulated BER results of the protograph codes. It reveals that the performance of DF protocol is approximately the same as that of EF protocol. Moreover, the theoretical BER expressions, which are shown to be reasonably consistent with the decoding thresholds and the simulated BERs, are able to evaluate the system performance and predict the decoding threshold with lower complexity as compared to the modified PEXIT algorithm. As a result, this work can facilitate the design of the protograph codes for the wireless communication systems.
\end{abstract}


\newpage

\section{Introduction}
Spatial diversity is an effective technique to enhance the quality and reliability of wireless communications and can be typically achieved using multiple antennas at a transmitter and/or a receiver. Besides, spatial diversity can also be obtained by the use of relaying. The relay channel, which consists of a source, a relay, and a destination, was probably first proposed in the 70s \cite{555172}. Later, Cover and Gamal have further developed the theory of such a channel \cite{1056084}. Recently, half-duplex relaying \cite{4107948} has been verified to be relatively more practical in comparison with the full-duplex one due to simpler implementation.

To improve the performance, forward error correction (FEC) codes have been applied to relay channels. As a capacity-approaching code, Low-density parity-check (LDPC) code has been developed to significantly enhance the performance of relay channels for different channel conditions, such as non-fading \cite{4107948,4305411,4686837} and fading scenarios \cite{4036109,5766201}. Moreover, \cite{4471943} provided the corresponding analysis of the LDPC-coded relay systems. At the same time, there has been a growing interest in protograph LDPC codes. It was shown in \cite{Thorpe2003ldp,4155107,5174517} that protograph (LDPC) codes not only achieve superior performance but also possess simple structures to realize linear encoding and decoding. For this reason, protograph codes have been used in additive white Gaussian noise (AWGN) relay channels \cite{5513451}, partial response channels \cite{6253209}, and Rayleigh fading channels \cite{6133952,Fang2012}. However, thus far, its analytical performance for general fading relay channels is not very well understood, which has motivated the results of this paper.

In this paper, we study the error performance of a half-duplex protograph LDPC-coded relay system over Nakagami-$m$ fading channels. The Nakagami-$m$ fading channel is a general type of fading channel that encompasses Rayleigh fading ($m=1$) and AWGN ($m \to \infty$) channels as special cases. We first use the modified protograph extrinsic information transfer (PEXIT) algorithm \cite{Fang2012} to analyze the decoding threshold of the protograph codes with different fading depths (i.e., different values of $m$). Afterwards, we derive the bit-error-rates (BERs) for the error-free (EF) and the decode-and-forward (DF) relaying protocols exploiting the modified PEXIT algorithm and using Gaussian approximation \cite{1246023}, and find that the BER analytical method has lower computational complexity than that of the modified PEXIT one. Simulated results show that the performance of the DF protocol approaches very close to that of EF, which align well with the analytical results. Furthermore, the results also indicate that the performance of the codes is improved as the fading depth decreases (higher $m$), but the rate of improvement is reduced simultaneously.

The remainder of this paper is organized as follows. In Section~\ref{sect:review}, the system model over the Nakagami-$m$ fading channel is described. In Section~\ref{sect:BER-Analysis}, we analyze the protograph LDPC coded relay system using the modified PEXIT algorithm. Moreover, we derive the BER expressions of the protograph codes. Numerical results are carried out in Section~\ref{sect:sim_dis}, and conclusions are given in Section~\ref{sect:conclusion}.

\section{System Model}\label{sect:review}
\begin{figure}[t]
\center
\includegraphics[width=3.2in,height=1.0in]{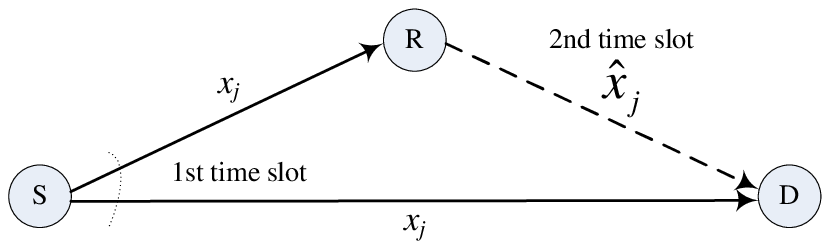}
\vspace{-0.3cm}
\caption{The system model of the protograph LDPC-coded relay system over fading channels.}\label{fig:Fig.1}
\end{figure}

We consider a two-hop half-duplex relay system model with one source, ${\sf S}$, one relay, ${\sf R}$, and one destination, ${\sf D}$, as shown in Fig.~\ref{fig:Fig.1}. Each transmission period is divided into two time slots, with the first slot being the {\em broadcast} time slot and the second slot being the {\em cooperative} time slot. During the first time slot, ${\sf S}$ broadcasts the codeword to other terminals (including ${\sf R}$ and ${\sf D}$). Then, in the second time slot, ${\sf R}$ cooperates with ${\sf S}$ to forward the re-encoded message of ${\sf S}$ to ${\sf D}$ while ${\sf S}$ remains idle.\footnote{In our model, we assume that the relay and the source adopt the same coding scheme.} During each transmission period, ${\sf D}$ stores the received codeword for decoding at the end of the second time slot. Mathematically, the received signals can be written as
\begin{align}
r_{{\rm R1},j} &=  h_{{\rm SR},j} x_j + n_{{\rm SR},j},\label{eq:SR-receiver}\\
r_{{\rm D1},j} &=  h_{{\rm SD},j} x_j + n_{{\rm SD},j},\label{eq:SD-receiver}\\
r_{{\rm D2},j} &=  h_{{\rm RD},j} \hat{x}_j + n_{{\rm RD},j},\label{eq:RD-receiver}
\end{align}
where, $x_j$ and $\hat{x}_j$ denote the binary-phase-shift-keying (BPSK) modulated signals corresponding to the $j$th coded bit $v_j$ and re-encoded bit $\hat{v}_j$ , respectively; $r_{{\rm R1},j}$, $r_{{\rm D1},j}$, $r_{{\rm D2},j}$ represent the received signals of the $j$th coded bit at the relay in the $1$st time slot, at the destination in the $1$st time slot, and at the destination in the $2$nd time slot, respectively; $h_{{\rm SR},j}$, $h_{{\rm SD},j}$, and $h_{{\rm RD},j}$ are the mutually independent Nakagami-$m$ fading channel coefficients of the S-R link, the S-D link, and the R-D link, respectively; and $n_{{\rm SR},j}$, $n_{{\rm SD},j}$, and $n_{{\rm RD},j}$ are the AWGN with zero mean and variance of $\sigma_n^2$
($\sigma_n^2=\frac{N_0}{2}$).

We define the channel gain of the S-R link as $\gamma_{{\rm SR},j}= |h_{{\rm SR},j}|^2$ (similar expressions will be assumed for other links). Moreover, we assume that the receiver (${\sf R}$ or ${\sf D}$) knows perfectly the channel state information (CSI) for decoding. At the destination, the signals from the source and relay are combined by a maximum ratio combiner (MRC). To simplify the analysis, we also assume that the distance from ${\sf S}$ to ${\sf D}$ is normalized to unity, while the distances from ${\sf S}$ to ${\sf R}$ and from ${\sf R}$ to ${\sf D}$ are $d$ and $1 - d$, respectively.\footnote{Note that the assumption of collinearity of ${\sf S}$, ${\sf R}$ and ${\sf D}$ will not affect any of the derivation of our results.} Based on this assumption, we have $h_{{\rm SD},j} = \alpha_{{\rm SD},j}$, $h_{{\rm SR},j}=\frac{\alpha_{{\rm SR},j}}{d}$, and $h_{{\rm RD},j}= \frac{\alpha_{{\rm RD},j}}{1-d}$, where $\alpha$ is the Nakagami-$m$ fading parameter with the probability density function (PDF) expressed as
\begin{equation}\label{eq:Nakagami_PDF}
f (x)= \frac{2} {\Gamma (m)} \left( \frac{m} {\Omega} \right)^{m} x^{2 m - 1}\exp \left( - \frac{m} {\Omega} x^2 \right),~~ \mbox{for }m \ge 0.5,
\end{equation}
where $m$ is the fading depth, $\Omega = {\mathbb E} [\alpha^2] = 1$, ${\mathbb E}[\cdot]$ is the expectation
operator, and $\Gamma(\cdot)$ is the Gamma function. In this paper, we focus on ergodic channels, in which the channel fades significantly rapidly such that it varies bit by bit.\footnote{Although we consider the ergodic (i.e., fast fading) scenario here, the results in this paper are also applicable for the quasi-static case (where the fading parameter of each link is kept to be constant for a code block).}

\section{Performance Analysis}\label{sect:BER-Analysis}
In this section, we derive the BER expression of the protograph code in our system with the EF and DF protocols based on the modified PEXIT algorithm and using Gaussian approximation \cite{1246023}. In the analysis, it is assumed that the all-zero codeword is transmitted and the block length of the code is infinite \cite{1291808}. The output (extrinsic and \textit{a-posteriori}) log-likelihood-ratio (LLR) messages of Nakagami-$m$ fading channels are approximately following the symmetric Gaussian distribution \cite{924876}.

Firstly, we review the protograph code \cite{Thorpe2003ldp}. A protograph is a Tanner graph with a relatively small number of nodes. A protograph code is a large protograph (called derived graph) obtained by the ``copy-and-permute'' operation on a protograph. Accordingly, a code with different block lengths can be produced by different number of times of the ``copy-and-permute" operation. A protograph with $N$ variable nodes and $M$ check nodes can be described by a base matrix $\bB = (b_{i,j})$ with dimensions $M\times N$, where $b_{i,j}$ denotes the number of edges connecting the variable node $v_j$ to the check node $c_i$. Since the protograph code always has some punctured variable nodes, we define $P_j$ as the punctured label of $v_j$ ($P_j=0$ if $v_j$ is punctured; otherwise $P_j=1$). We also define the following LLRs and mutual information (MI).

{\bf LLR value:}
\begin{itemize}
\item
$L_{\rm Av}(i,j)$ denotes the input (\textit{a-priori}) LLR value of $v_j$ corresponding to $c_i$ on each of the $b_{i,j}$ edges.
\item
$L_{\rm Ac}(i,j)$ denotes the input (\textit{a-priori}) LLR value of $c_i$ corresponding to $v_j$ on each of the $b_{i,j}$ edges.
\item
$L_{\rm Ev}(i,j)$ denotes the output (extrinsic) LLR value passing from $v_j$ to the $c_i$.
\item
$L_{\rm Ec}(i,j)$ denotes the output (extrinsic) LLR value passing from $c_i$ to the $v_j$.
\item
$L_{\rm app}(j)$ denotes the \textit{a-posteriori} LLR value of $v_j$.
\end{itemize}

{\bf MI:}
\begin{itemize}
\item $I_{\rm Av}(i, j)$ denotes the \textit{a-priori} MI between $L_{\rm Av}(i,j)$ and the corresponding coded bit $v_j$.
\item $I_{\rm Ac}(i, j)$ denotes the \textit{a-priori} MI between $L_{\rm Ac}(i,j)$ and the corresponding coded bit $v_j$.
\item $I_{\rm Ev}(i,j)$ denotes the extrinsic MI between $L_{\rm Ev}(i,j)$ and the corresponding coded bit $v_j$.
\item $I_{\rm Ec}(i, j)$ denotes the extrinsic MI between $L_{\rm Ec}(i,j)$ and the corresponding coded bit $v_j$.
\item $I_{\rm app}(j)$ denotes the \textit{a-posteriori} MI between $L_{\rm app}(j)$ and the corresponding coded bit $v_j$.
\end{itemize}

Note that the subscripts ``SR'', ``SD", ``RD'', and ``D"   are used to denote the S-R link, the S-D link, the R-D link, and the destination, respectively. For example, $L_{\rm SR-Av}(i,j)$ represents the a-priori LLR value of $v_j$ corresponding to $c_i$ on each of the $b_{i,j}$ edges of the S-R link.

\subsection{BER of EF Relaying Protocol}\label{sect:EF-BER}
At the destination receiver, the output of MRC corresponding to the $j$th coded bit is given by
\begin{equation}\label{eq:MRC-output}
y_{{\rm D},j} = h_{{\rm SD},j} r_{{\rm D1},j} +  h_{{\rm RD},j} r_{{\rm D2},j}.
\end{equation}
Afterwards, the initial LLR value $L_{{\rm D-ch},j}$ of the $j$th coded bit is calculated as
\begin{align}
L_{{\rm D-ch},j} &= \displaystyle \ln \frac{ \Pr(v_j=0|y_{{\rm D},j}, \bh_j)}{ \Pr(v_j=1|y_{{\rm D},j}, \bh_j)} = \ln \frac{ \Pr(x_j=+1|y_{{\rm D},j}, \bh_j)}{ \Pr(x_j=-1|y_{{\rm D},j}, \bh_j)} \notag\\
&= \frac{2 y_{{\rm D},j}} {\sigma_n^2} = \frac{2} {\sigma_n^2} (h_{{\rm SD},j} r_{{\rm D1},j} +  h_{{\rm RD},j} r_{{\rm D2},j})\notag\\
&= \frac{2} {\sigma_n^2} \left[ h_{{\rm SD},j} (h_{{\rm SD},j} x_j + n_{{\rm SD},j}) + h_{{\rm RD},j} (h_{{\rm RD},j} \hat{x}_j + n_{{\rm RD},j}) \right] \notag\\
&=  \frac{2} {\sigma_n^2} \left[ |h_{{\rm SD},j}|^2  x_j  + |h_{{\rm RD},j}|^2 \hat{x}_j + h_{{\rm SD},j} n_{{\rm SD},j} + h_{{\rm RD},j} n_{{\rm RD},j} \right] \notag\\
&=  \frac{2} {\sigma_n^2} \left[ (\gamma_{{\rm SD},j} + \gamma_{{\rm RD},j}) x_j + h_{{\rm SD},j} n_{{\rm SD},j} + h_{{\rm RD},j} n_{{\rm RD},j} \right],\label{eq:L-D-ch}
\end{align}
where $\Pr(\cdot)$ denotes the probability function, $\bh_j = \{h_{{\rm SR},j}, h_{{\rm SD},j}, h_{{\rm RD},j} \}$ is the channel vector, $\gamma_{{\rm SD},j}=|h_{{\rm SD},j}|^2=\alpha_{{\rm SD},j}^2$, $\gamma_{{\rm RD},j}=|h_{{\rm RD},j}|^2=\frac{\alpha_{{\rm RD},j}^2}{(1-d)^2}$, and $x_j=\hat{x}_j$ (error-free decoding at the relay).

Based on the all-zero codeword assumption ($x_j = +1$), given a fixed channel realization (a fixed fading vector $\bh_j $), we can easily obtain the expectation and the variance of $L_{{\rm D-ch},j}$, resulting in
\begin{align}
{\mathbb E} [L_{{\rm D-ch},j}] &=\frac{2} {\sigma_n^2} (\gamma_{{\rm SD},j} + \gamma_{{\rm RD},j}) = \frac{2} {\sigma_n^2} \lambda_{{\rm D},j}, \label{eq:mean-L-D-ch}\\
{\rm var} [L_{{\rm D-ch},j}] &=\frac{4} {\sigma_n^4} (\gamma_{{\rm SD},j} \sigma_n^2 + \gamma_{{\rm RD},j} \sigma_n^2) = \frac{4} {\sigma_n^2} \lambda_{{\rm D},j},\label{eq:var-L-D-ch}
\end{align}
where $\lambda_{{\rm D},j}$ is the short-hand notation of $\gamma_{{\rm SD},j} + \gamma_{{\rm RD},j}$. As seen from \eqref{eq:mean-L-D-ch} and \eqref{eq:var-L-D-ch}, $L_{{\rm D-ch},j}$ follows the symmetric Gaussian distribution, i.e., $L_{{\rm D-ch},j} \sim {\cal N} (\frac{2} {\sigma_n^2} \lambda_{{\rm D},j}, \frac{4} {\sigma_n^2} \lambda_{{\rm D},j})$. Subsequently, substituting $\sigma_n^2=\frac{1} {2R(E_b'/N_0)}$ into \eqref{eq:var-L-D-ch} ($E_b'=\frac{E_b}{2}$, the normalized factor $1/2$ is used to keep the total energy per transmitted symbol to be $E_s$) and considering the punctured label $P_j$, we have
\begin{equation}\label{eq:simplied-var-L-D-ch}
{\rm var} [L_{{\rm D-ch},j}] =  8 R P_j \lambda_{{\rm D},j} (E_b'/N_0) = 4 R P_j \lambda_{{\rm D},j} (E_b/N_0).
\end{equation}
Thus, the modified PEXIT algorithm \cite{Fang2012} can be applied to our system by using the expression \eqref{eq:simplied-var-L-D-ch}.\footnote{According to \cite{Fang2012}, the maximum iteration number of the modified PEXIT algorithm $T_{\rm max}^{\rm P}$ should be large enough in order to ensure the complete convergence of the decoder, i.e., $T_{\rm max}^{\rm P}\ge 500$.}

For the $q$th ($q=1,2\ldots,Q$, where $Q$ is the total number of channel realizations\footnote{To ensure the accuracy of our derived BER expressions, $Q$ should be set to a sufficiently large integer, i.e., $Q \ge 10^5$.}) channel realization $h_{q,j}$, the a-posteriori LLR of $v_j$ during the $(t+1)$th iteration is written as
\begin{equation}\label{eq:L-D-app}
L_{{\rm D-app},q}^{t+1} (j) =  \sum_{i=1}^{M} b_{i,j} L_{{\rm D-Av},q}^{t+1} (i,j) + L_{{\rm D-ch},q,j}.
\end{equation}
The expected value of $L_{{\rm D-Av},q}^{t+1} (i,j)$ is then expressed by
\begin{equation}\label{eq:simplied-L-D-app}
\bar{L}_{{\rm D-app}}^{t+1} (j) =  {\mathbb E} [L_{{\rm D-app},q}^{t+1} (j)] = \frac{1} {Q} \sum_{q=1}^{Q} L_{{\rm D-app},q}^{t+1} (j).
\end{equation}
As the output LLR values, i.e., the extrinsic LLR and a-posteriori LLR, are approximated to follow a symmetric Gaussian distribution, we can evaluate the variance of $\bar{L}_{{\rm D-app}}^{t+1} (j)$ as
\begin{equation}\label{eq:var-L-D-app}
{\rm var} [\bar{L}_{{\rm D-app}}^{t+1} (j)] =  \left\{ J^{-1} (\bar{I}_{{\rm D-app}}^{t+1} (j)) \right\}^2 = \left\{ J^{-1} ({\mathbb E} [I_{{\rm D-app},q}^{t+1} (j)]) \right\}^2,
\end{equation}
where $\bar{I}_{{\rm D-app}}^{t+1} (j)={\mathbb E} [I_{{\rm D-app},q}^{t+1} (j)]=\frac{1}{Q} \sum_{q=1}^{Q} I_{{\rm D-app},q}^{t+1} (j)$ is the expected value of $I_{{\rm D-app},q}^{t+1} (j)$ over all channel realizations, the expression of the a-posteriori MI $I_{{\rm D-app},q}^{t+1} (j)$ is obtained as \cite[(21)]{Fang2012}, denoted by $I_{{\rm D-app},q}^{t+1} (j)=F(I_{{\rm D-Av},q}^{t+1} (1,j),\ldots,I_{{\rm D-Av},q}^{t+1} (M,j);{\rm var} [L_{{\rm D-ch},q,j}])$, and $J(\sigma_{\rm ch})$ is the MI between a BPSK modulated bit and its LLR value $L_{\rm ch} \sim {\cal N} (\frac{\sigma_{\rm ch}^2}{2}, \sigma_{\rm ch}^2)$ over an AWGN channel, represented as \cite{1291808}
\begin{equation}\label{eq:J-fuction}
J(\sigma_{\rm ch}) = 1- \int_{-\infty}^{\infty} \frac{\exp \left( -  \frac{(\xi -\sigma_{\rm ch}^2/2)^2 }{2 \sigma_{\rm ch}^2} \right)}{\sqrt{2\pi \sigma_{\rm ch}^2}}
\log_2 \left[ 1+\exp(-\xi) \right] {\rm d} \xi .
\end{equation}
The corresponding inverse function is given by \cite{1291808}
\begin{equation}\label{eq:Inverse-J-fuction}
J^{-1}(x) =\left\{\begin{array}{ll}
 \eta_1 x^2 + \eta_2 x + \eta_3 \sqrt{x} & \;\; \;\;  \text{if} \;\; 0 \le x \le 0.3646,  \\
\eta_4 \ln[\eta_5(1-x)] + \eta_6 x & \;\;\;\;   \text{otherwise},
\end{array}\right.
\end{equation}
where $\eta_1=1.09542$, $\eta_2=0.214217$, $\eta_3=2.33737$, $\eta_4= -0.706692$, $\eta_5=0.386013$ and $\eta_6=1.75017$. With the help of \eqref{eq:var-L-D-app}--\eqref{eq:Inverse-J-fuction}, the BER of the $j$th variable node after $t$ iterations is evaluated by
\begin{align}
P_{{\rm D}-b}^{t+1} (j) &= \frac{1} {2} \Pr(\bar{L}_{{\rm D-app}}^{t+1} (j)<0|x_j=+1) +
\frac{1} {2} \Pr(\bar{L}_{{\rm D-app}}^{t+1} (j) \ge 0|x_j=-1)\notag\\
&= \frac{1} {2} \text{erfc} \left(\frac{{\mathbb E}[\bar{L}_{{\rm D-app}}^{t+1} (j)|x_j=+1]} {\sqrt{2 {\rm var}[\bar{L}_{{\rm D-app}}^{t+1} (j)|x_j=+1]}} \right) = \displaystyle \frac{1} {2} \text{erfc} \left(\frac{{\rm var}[\bar{L}_{{\rm D-app}}^{t+1} (j)|x_j=+1]/2} {\sqrt{2 {\rm var}[\bar{L}_{{\rm D-app}}^{t+1} (j)|x_j=+1]}} \right)\notag\\
&= \frac{1} {2} \text{erfc} \left(\frac{\sqrt{{\rm var}[\bar{L}_{{\rm D-app}}^{t+1} (j)|x_j=+1]}} {2 \sqrt{2}} \right)= \frac{1} {2} \text{erfc} \left(\frac{ J^{-1} (\bar{I}_{{\rm D-app}}^{t+1} (j))} {2 \sqrt{2}} \right),\label{eq:Pb-D-bit}
\end{align}
where $\text{erfc}(\cdot)$ is the complementary error function, defined as $\text{erfc} (x) =\frac{2}  {\sqrt{\pi}} \int_{x}^{\infty} e^{-\tau^2} {\rm d}\tau$.

Finally, the averaged BER of a protograph code after $t$ iterations is written as
\begin{equation}\label{eq:Pb-D-average}
P_{{\rm EF}-b}^{t+1} = P_{{\rm D}-b}^{t+1} = \frac{1} {N} \sum_{j=1}^{N} P_{{\rm D}-b}^{t+1} (j)
=  \frac{1} {2 N} \sum_{j=1}^{N}  \text{erfc} \left(\frac{ J^{-1} (\bar{I}_{{\rm D-app}}^{t+1} (j))} {2 \sqrt{2}} \right).
\end{equation}

\subsection{BER of DF Relaying Protocol}\label{sect:DF-BER}
In DF, if the relay can decode received signals correctly, it re-encodes the decoded information and then forwards them to the destination; otherwise, it does not send message or remains idle. Consequently, the BER of the DF protocol of the $j$th variable node after $t$ iterations can be found as
\begin{equation}\label{eq:Pb-DF-bit}
P_{{\rm DF}-b}^{t+1} (j) = P_{{\rm SR}-b}^{t+1} (j) P_{{\rm SD}-b}^{t+1} (j) + \left[ 1 - P_{{\rm SR}-b}^{t+1} (j) \right] P_{{\rm D}-b}^{t+1} (j),
\end{equation}
where $P_{{\rm SR}-b}^{t+1} (j)$, $P_{{\rm SD}-b}^{t+1} (j)$, and $P_{{\rm D}-b}^{t+1} (j)$ are the corresponding BERs  of the $j$th variable node after $t$ iterations at the relay receiver with the signal from source, at the destination receiver with one signal from the source, and at the destination receiver with two signals both from the source and relay (i.e., BER of the EF protocol), respectively.

 The initial LLR of the $j$th variable node of the S-R link is expressed by \cite{924876}
\begin{equation}\label{eq:L-SR-ch}
L_{{\rm SR-ch},j} = \frac{2 h_{{\rm SR},j} r_{{\rm R1},j}} {\sigma_n^2} = \frac{2 h_{{\rm SR},j}} {\sigma_n^2} (h_{{\rm SR},j} x_j + n_{{\rm SR},j}) = \frac{2} {\sigma_n^2} ( \gamma_{{\rm SR},j} x_j + h_{{\rm SR},j} n_{{\rm SR},j}).
\end{equation}
In \eqref{eq:L-SR-ch}, we have ${\mathbb E} [L_{{\rm SR-ch},j}] = \frac{2 \gamma_{{\rm SR},j}} {\sigma_n^2}$ and ${\rm var} [L_{{\rm SR-ch},j}] =
\frac{4 \gamma_{{\rm SR},j}} {\sigma_n^2}$. Hence, during the $(t+1)$th iteration, the a-posterior MI of the $j$th variable for the $q$th ($q=1,2,\ldots,Q$)
channel realization and the corresponding expected value are respectively given by 
\begin{align}
I_{{\rm SR-app},q}^{t+1} (j) &=F(I_{{\rm SR-Av},q}^{t+1} (1,j),\ldots,I_{{\rm SR-Av},q}^{t+1} (M,j);{\rm var} [L_{{\rm SR-ch},q,j}]), \label{eq:I-SR-app}\\
\bar{I}_{{\rm SR-app}}^{t+1} (j) &= {\mathbb E} [I_{{\rm SR-app},q}^{t+1} (j)] = \frac{1} {Q} \sum_{q=1}^{Q} I_{{\rm SR-app},q}^{t+1} (j).\label{eq:avrage-I-SR-app}
\end{align}

Further, the variance of the expected a-posterior LLR associated with \eqref{eq:avrage-I-SR-app} is yielded in terms of the Gaussian assumption
\begin{equation}\label{eq:var-L-SR-app}
{\rm var} [\bar{L}_{{\rm SR-app}}^{t+1} (j)] =  \left\{ J^{-1} (\bar{I}_{{\rm SR-app}}^{t+1} (j)) \right\}^2.
\end{equation}
Note that $ {\mathbb E} [\bar{L}_{{\rm SR-app}}^{t+1} (j)] = {\rm var} [\bar{L}_{{\rm SR-app}}^{t+1} (j)]/2$.

Using \eqref{eq:var-L-SR-app}, we can therefore obtain the BER of the $j$th variable after $t$ iterations as
\begin{equation}\label{eq:Pb-SR-bit}
P_{{\rm SR}-b}^{t+1} (j) = \frac{1} {2} \text{erfc} \left(\frac{{\mathbb E}[\bar{L}_{{\rm SR-app}}^{t+1} (j)|x_j=+1]} {\sqrt{2 {\rm var}[\bar{L}_{{\rm SR-app}}^{t+1} (j)|x_j=+1]}} \right) = \displaystyle \frac{1} {2} \text{erfc} \left(\frac{ J^{-1} (\bar{I}_{{\rm SR-app}}^{t+1} (j))} {2 \sqrt{2}} \right).
\end{equation}

Likewise, the BER of the $j$th variable after $t$ iterations of the S-D link can be written as
\begin{equation}\label{eq:Pb-SD-bit}
P_{{\rm SD}-b}^{t+1} (j) = \frac{1} {2} \text{erfc} \left(\frac{ J^{-1} (\bar{I}_{{\rm SD-app}}^{t+1} (j))} {2 \sqrt{2}} \right)
\end{equation}
with the parameters subjected to
\begin{equation}\label{eq:Pb-SD-bit-para}
\left\{\begin{aligned}
&\bar{I}_{{\rm SD-app}}^{t+1} (j) =  {\mathbb E} [I_{{\rm SD-app},q}^{t+1} (j)] = \frac{1} {Q} \sum_{q=1}^{Q} I_{{\rm SD-app},q}^{t+1} (j)\\
&I_{{\rm SD-app},q}^{t+1} (j) = F(I_{{\rm SD-Av},q}^{t+1} (1,j),\ldots,I_{{\rm SD-Av},q}^{t+1} (M,j);{\rm var} [L_{{\rm SD-ch},q,j}])\\
&{\rm var} [L_{{\rm SD-ch},q,j}] = \frac{4 \gamma_{{\rm SD},q,j}} {\sigma_n^2},
\end{aligned}\right.
\end{equation}
where $L_{{\rm SD-ch},q,j}$ is the initial LLR of $v_j$ for the $q$th S-D link realization and $\gamma_{{\rm SD},q,j}=|h_{{\rm SD},q,j}|^2$.

Substituting \eqref{eq:Pb-D-bit}, \eqref{eq:Pb-SR-bit} and \eqref{eq:Pb-SD-bit} into \eqref{eq:Pb-DF-bit}, the BER of the $j$th variable node with DF after $t$ iterations can be formulated. Finally, we get the averaged BER with DF protocol after $t$ iterations as
\begin{equation}\label{eq:Pb-DF-average}
P_{{\rm DF}-b}^{t+1} = \frac{1} {N} \sum_{j=1}^{N} P_{{\rm DF}-b}^{t+1} (j).
\end{equation}

Note also that
\begin{itemize}
\item The maximum number of iterations of the theoretical BER analysis $T_{\rm max}$ ($t=T_{\rm max}$) equals to that of the simulations ($T_{\rm max}$ is always much smaller than $T_{\rm max}^{\rm P}$). In this paper, we set $T_{\rm max}=100$ as in \cite{Fang2012,5751586}. Therefore, its computational complexity can be reduced by $\frac{(T_{\rm max}^{\rm P}-T_{\rm max})} {T_{\rm max}^{\rm P}}$ as compared to the modified PEXIT algorithm.
\item The BER analysis can be used to evaluate the performance of the protograph codes for any $E_b/N_0$ and $T_{\rm max}$, while the modified PEXIT algorithm can only be exploited to derive the $E_b/N_0$ threshold above which an arbitrarily small BER can be achieved (i.e., all the code blocks can be successfully decoded) for a sufficiently large ${T_{\rm max}^{\rm P}}$.
\end{itemize}

\section{Numerical Results and Discussions}\label{sect:sim_dis}
In this section, we firstly analyze the decoding threshold of two typical protograph codes, i.e., the AR3A and AR4JA codes, over Nakagami-$m$ fading relay channels utilizing the modified PEXIT algorithm \cite{Fang2012}. Then, we compare the decoding thresholds, the theoretical BER and simulated BER results of the two protograph codes. For all the following results, the distance of S-R link $d$ is set to $0.4$.

The AR3A and AR4JA codes, which have been proposed by Jet Propulsion Laboratory \cite{4155107,5174517}, can respectively accomplish excellent performance in the low SNR region and the high SNR region over the AWGN channel. The corresponding base matrices of these two codes with a code rate of $R =\frac{n+1}{n+2}$, denoted by $\bB_{A3}$ and $\bB_{A4}$, respectively, are given by
\begin{align}
\bB_{A3} &=\left(\begin{array}{llllll}
1 & 2 & 1 & 0 & 0  & \overbrace{0 \ 0 \ \cdots \ 0 \ 0}^{2n} \cr
0 & 2 & 1 & 1 & 1  & 2 \ 1 \ \cdots \ 2 \ 1 \cr
0 & 1 & 2 & 1 & 1  & 1 \ 2 \ \cdots \ 1 \ 2
 \end{array}\right),\label{eq:BA3}\\
\bB_{A4} &=\left(\begin{array}{llllll}
1 & 2 & 0 & 0 & 0  & \overbrace{0 \ 0 \ \cdots \ 0 \ 0}^{2n} \cr
0 & 3 & 1 & 1 & 1  & 3 \ 1 \ \cdots \ 3 \ 1 \cr
0 & 1 & 2 & 2 & 1  & 1 \ 3 \ \cdots \ 1 \ 3
 \end{array}\right).\label{eq:BA4}
\end{align}
In \eqref{eq:BA3} and \eqref{eq:BA4}, the $j$th column corresponds to the $j$th variable node and the $i$th row refers to the $i$th check node. The variable nodes corresponding to the second columns of the two matrices are punctured.

\subsection{Decoding Threshold Analysis}
We calculate the decoding thresholds of the AR3A and AR4JA codes with different code rates and different fading depths using the modified PEXIT algorithm \cite{Fang2012} and show the result in Table~\ref{tab:Thre-diff-m}.\footnote{For the threshold analysis, we only consider the perfect S-R channel (i.e., EF), as is typical in many practical cases \cite{969907}.} As seen from this table, the threshold of the AR3A code is lower than that of the AR4JA code for a fixed code rate and fading depth (fixed $R$ and $m$). For instance, the thresholds of the AR3A code and the AR4JA code are $0.575$~dB and $0.722$~dB, respectively, with parameters $R=4/5$ and $m=2$. Moreover, the threshold is reduced as the fading depth decreases (higher $m$) for both the two codes and hence the error performance should be improved in the waterfall region. However, the decrease of the threshold is reduced when $m$ becomes larger, indicating that the rate of performance improvement is reduced.

\subsection{BER Performance}
In the sequel, we present the simulated results of the AR3A and AR4JA codes with a code rate of $R = 4/5$ and compare their simulated and theoretical BER curves with the decoding thresholds. We denote $N_{\rm p}$, $K_{\rm p}$, and $L_{\rm p}$ as the block length, the information length, and the punctured length of the code, respectively. Unless specified otherwise, in the simulations, it has been assumed that the decoder performs a maximum of $100$ iterations for each code block with parameters $[N_{\rm p}, K_{\rm p}, L_{\rm p}] = [5632, 4096, 512]$.

\begin{figure}[tbp]
\center
\includegraphics[width=3.0in,height=2.3in]{{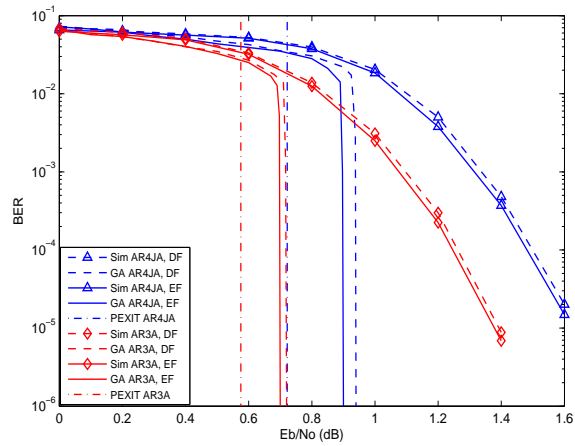}}
\vspace{-0.2cm}
\caption{BER results of AR3A and AR4JA codes over Nakagami-fading relay channels with the fading depth $m=2$.}
\label{fig:Fig.2}  
\end{figure}

In Fig.~\ref{fig:Fig.2}, we show the BER results (vs.~$E_b/N_0$) of the AR4JA and AR3A
codes over Nakagami-fading relay channels with DF and EF protocols. The fading depth $m$
is set to $2$. As seen, the error performance of the codes with the DF protocol approaches
closely to that with EF, which suggests that the relay can decode most of the received
code blocks successfully. Moreover, the AR3A code outperforms the AR4JA codes for the
range of $E_b/N_0$ under study. At a BER of $10^{-5}$, the AR3A code achieves
a gain approximately $0.2$~dB as compared to the AR4JA code both for the DF and EF
protocols. In the same figure, consider the AR3A code with DF protocol at a BER of
$10^{-5}$, the theoretical BER result is in good agreement with the PEXIT threshold
within $0.2$~dB. Moreover, the corresponding simulated curve has another gap about $0.7$~dB
to the theoretical one, which is reasonably consistent with the results in \cite{5751586,1246023}. It is because our theoretical BER formulas are derived
subjecting to the infinite-block length assumption.

\begin{figure}[tbp]
\centering
\includegraphics[width=3.0in,height=2.3in]{{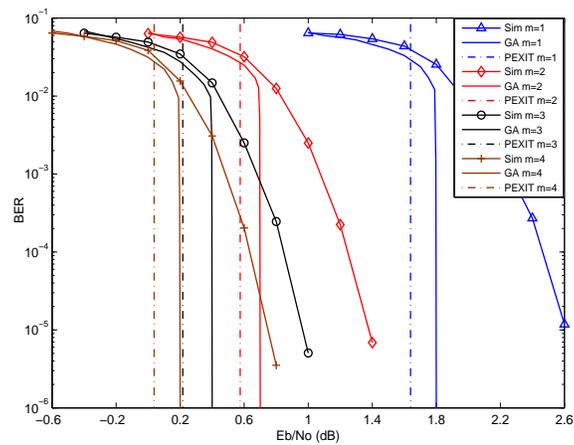}}
\vspace{-0.2cm}
\caption{BER results of the AR3A code over Nakagami-fading relay channels with different fading depths ($m=1,2,3,$ and $4$).}
\label{fig:Fig.3}  
\end{figure}

Fig.~\ref{fig:Fig.3} presents the BER results (vs.~$E_b/N_0$) of the AR3A code in the EF relay systems for different fading depths ($m = 1,2,3,$ and $4$). As $E_b/N_0$ increases steadily, both the two codes start to perform better for a larger value of $m$ in terms of the thresholds, the theoretical and simulated BER curves. Nevertheless, the improved gain is reduced as $m$ increases. For example, at a BER of $10^{-5}$, the improved gains are about $1.2$~dB, $0.45$~dB, and $0.25$~dB as $m$ increases from $1$ to $2$, $2$ to $3$, and $3$ to $4$, respectively. Moreover, it can be observed that gaps between the theoretical BERs and the simulated results are around $0.8$~dB, $0.7$~dB, $0.55$~dB, and $0.5$~dB for $m = 1,2,3,$ and $4$, respectively. Simulations have also been performed for the AR4JA code and with the DF protocol, and similar observations are obtained.

\section{Conclusions}\label{sect:conclusion}
The performance of the protograph LDPC codes over Nakagami-$m$ fading relay channels has been studied. The BER expressions for the protograph codes with DF and EF relaying protocols have been derived using the modified PEXIT algorithm and Gaussian approximation. The decoding threshold, the theoretical BER and the simulated BER results have shown that the error performance of the DF protocol is very close to that of EF, which suggests that the relay can decode most received codewords correctly. The differences between the threshold and the theoretical BER and between the theoretical BER and the simulated BER have been found to be around $0.15-0.2$~dB and $0.5-0.8$~dB, respectively, showing a reasonable consistence. Consequently, our analytical expressions not only can provide a good approximation of the system performance for a large-block length but also predict accurately the decoding threshold more efficiently in comparison with the modified PEXIT algorithm.

\section{Acknowledgments}
The authors would like to thank the anonymous reviewers for
their valuable comments that have helped improving the
overall quality of the paper. This work was partially
supported by the NSF of China (nos. 61271241,
61001073 and 61102134), the European Union-FP7
(CoNHealth, no. 294923), as well as the Fundamental Research Funds for the Central Universities (No. 201112G017).


\newpage

\listoffigures

\listoftables

\newpage

\begin{table*}[htbp]
\caption{Decoding thresholds $(E_b/N_0)_{\rm th}$~({\rm dB}) of the AR3A code and AR4JA code with different code rates over Nakagami-fading relay channels with different fading depths $m =1$, $2$, $3$, and $4$.}
\centering
\begin{tabular}{|c|c|c|c|c|c|c|c|c|}
    \hline
    \multirow{2}{*}{Code Rate}
    & \multicolumn{4}{|c|}{AR3A code} & \multicolumn{4}{|c|}{AR4JA code}\\
    \cline{2-9}
    & $m = 1$ &$m = 2$ & $m = 3$ & $m = 4$
    & $m = 1$ &$m = 2$ & $m = 3$ & $m = 4$ \\
    \hline
    $1/2 \, (n=0)$ & $-1.345$ &	$-1.825$ & $-1.983$ & $-2.056$ & $-1.187$ & $-1.675$ &	
    $-1.828$ & $-1.895$\\
    \hline
    $2/3 \, (n =1)$ & $0.004$ &	$-0.759$ & $-0.994$ & $-1.098$ & $0.162$ & $-0.566$ &
    $-0.823$ & $-0.918$\\
    \hline
    $3/4 \, (n =2)$ & $0.890$ &	$-0.016$ &	$-0.316$ & $-0.472$ & $1.128$ & $0.185$ &
    $-0.136$ & $-0.292$\\
    \hline
    $4/5 \, (n =3)$ & $1.639$ & $0.575$	& $0.216$ & $0.037$ & $1.805$ &	$0.722$	&
    $0.382$	& $0.192$\\
    \hline
    $5/6 \, (n =4)$ & $2.235$ & $1.042$ & $0.653$ & $0.454$ & $2.392$ &	$1.164$ &
    $0.785$ & $0.576$\\
    \hline
    $6/7 \, (n =5)$ & $2.732$ & $1.404$ & $0.991$ & $0.778$	& $2.855$ & $1.533$	&
    $1.092$	& $0.887$\\
    \hline
    $7/8 \, (n =6)$ & $3.158$ &	$1.748$	& $1.278$ & $1.044$	& $3.284$ &	$1.852$	&
    $1.379$	& $1.150$\\
    \hline
\end{tabular}
\label{tab:Thre-diff-m}
\end{table*}


\begin{thebibliography}{10}
\providecommand{\url}[1]{#1}
\csname url@samestyle\endcsname
\providecommand{\newblock}{\relax}
\providecommand{\bibinfo}[2]{#2}
\providecommand{\BIBentrySTDinterwordspacing}{\spaceskip=0pt\relax}
\providecommand{\BIBentryALTinterwordstretchfactor}{4}
\providecommand{\BIBentryALTinterwordspacing}{\spaceskip=\fontdimen2\font plus
\BIBentryALTinterwordstretchfactor\fontdimen3\font minus
  \fontdimen4\font\relax}
\providecommand{\BIBforeignlanguage}[2]{{%
\expandafter\ifx\csname l@#1\endcsname\relax
\typeout{** WARNING: IEEEtran.bst: No hyphenation pattern has been}%
\typeout{** loaded for the language `#1'. Using the pattern for}%
\typeout{** the default language instead.}%
\else
\language=\csname l@#1\endcsname
\fi
#2}}
\providecommand{\BIBdecl}{\relax}
\BIBdecl

\bibitem{555172}
E.~C. van~der Meulen, ``{Three-terminal communication channels},'' \emph{Adv.
  Appl. Probab.}, vol.~3, pp. 120--154, 1971.

\bibitem{1056084}
T.~Cover and A.~Gamal, ``{Capacity theorems for the relay channel},''
  \emph{IEEE Trans. Inf. Theory}, vol.~25, no.~5, pp. 572--584, Sep. 1979.

\bibitem{4107948}
A.~Chakrabarti, A.~de~Baynast, A.~Sabharwal, and B.~Aazhang, ``{Low density
  parity check codes for the relay channel},'' \emph{IEEE J. Sel. Areas
  Commun.}, vol.~25, no.~2, pp. 280--291, Feb. 2007.

\bibitem{4305411}
P.~Razaghi and W.~Yu, ``{Bilayer low-density parity-check codes for
  decode-and-forward in relay channels},'' \emph{IEEE Trans. Inf. Theory},
  vol.~53, no.~10, pp. 3723--3739, Oct. 2007.

\bibitem{4686837}
C.~Li, G.~Yue, X.~Wang, and M.~Khojastepour, ``{LDPC code design for
  half-duplex cooperative relay},'' \emph{IEEE Trans. Wireless Commun.},
  vol.~7, no.~11, pp. 4558--4567, Nov. 2008.

\bibitem{4036109}
J.~Hu and T.~Duman, ``{Low density parity check codes over half-duplex relay
  channels},'' in \emph{Proc. 2006 IEEE Int. Symp. Inform. Theory (ISIT'06)},
  Jul. 2006, pp. 972--976.

\bibitem{5766201}
O.~Vahabzadeh and M.~Salehi, ``{Design of bilayer lengthened LDPC codes for
  Rayleigh fading relay channels},'' in \emph{Proc. 2011 Annual Conf. Inf. Sci.
  and Syst. (CISS),}, Mar. 2011, pp. 1--5.

\bibitem{4471943}
C.~Li, G.~Yue, M.~Khojastepour, X.~Wang, and M.~Madihian, ``{LDPC-coded
  cooperative relay systems: performance analysis and code design},''
  \emph{IEEE Trans. Commun.}, vol.~56, no.~3, pp. 485--496, Mar. 2008.

\bibitem{Thorpe2003ldp}
J.~Thorpe, ``{Low-density parity-check (LDPC) codes constructed from
  protographs},'' in \emph{Proc. IPN Progress Report}, 2003, pp. 42--154.

\bibitem{4155107}
A.~Abbasfar, D.~Divsalar, and K.~Yao, ``{Accumulate-repeat-accumulate codes},''
  \emph{IEEE Trans. Commun.}, vol.~55, no.~4, pp. 692--702, Apr. 2007.

\bibitem{5174517}
D.~Divsalar, S.~Dolinar, C.~Jones, and K.~Andrews, ``{Capacity-approaching
  protograph codes},'' \emph{IEEE J. Sel. Areas Commun.}, vol.~27, no.~6, pp.
  876--888, Aug. 2009.

\bibitem{5513451}
T.~V. Nguyen, A.~Nosratinia, and D.~Divsalar, ``{Bilayer protograph codes for
  half-duplex relay channels},'' in \emph{Proc. 2010 IEEE Int. Symp. Inform.
  Theory (ISIT'10)}, Jun. 2010, pp. 948--952.

\bibitem{6253209}
Y.~Fang, P.~Chen, L.~Wang, and F.~C.~M. Lau, ``{Design of protograph LDPC codes for partial response channels},'' \emph{IEEE Trans.
  Commun.}, vol.~60, no.~10, pp. 2809-2819, Oct. 2012.

\bibitem{6133952}
T.~V. Nguyen, A.~Nosratinia, and D.~Divsalar, ``{Threshold of protograph-based
  LDPC coded BICM for Rayleigh fading},'' in \emph{Proc. IEEE Global Telecommun.
  Conf.}, Dec. 2011, pp. 1--5.

\bibitem{Fang2012}
Y.~Fang, P.~Chen, L.~Wang, F.~C.~M. Lau, and K.~K. Wong, ``{Performance
  analysis of protograph-based LDPC codes with spatial diversity},'' \emph{IET
  Commun.}, vol.~6, no.~17, pp. 2941--2948, Dec. 2012.

\bibitem{1246023}
F.~Lehmann and G.~Maggio, ``{Analysis of the iterative decoding of LDPC and
  product codes using the Gaussian approximation},'' \emph{IEEE Trans. Inf.
  Theory}, vol.~49, no.~11, pp. 2993--3000, Nov. 2003.

\bibitem{1291808}
S.~ten Brink, G.~Kramer, and A.~Ashikhmin, ``{Design of low-density
  parity-check codes for modulation and detection},'' \emph{IEEE Trans.
  Commun.}, vol.~52, no.~4, pp. 670--678, Apr. 2004.

\bibitem{924876}
J.~Hou, P.~Siegel, and L.~Milstein, ``{Performance analysis and code
  optimization of low density parity-check codes on Rayleigh fading
  channels},'' \emph{IEEE J. Sel. Areas Commun.}, vol.~19, no.~5, pp. 924
  --934, May 2001.

\bibitem{969907}
D.~Duyck, D.~Capirone, J.~J. Boutros, and M.~Moeneclaey,
  ``\BIBforeignlanguage{eng}{{Analysis and construction of full-diversity joint
  network-LDPC codes for cooperative communications}},''
  \emph{\BIBforeignlanguage{eng}{Eurasip J. Wireless Commun. Netw.}}, pp.
  1--16, Jan. 2010.

\bibitem{5751586}
B.~S.~Tan, K.~H.~Li and K.~C.~Teh,
  ``\BIBforeignlanguage{eng}{{Performance analysis of LDPC codes with maximum-ratio combining cascaded with selection combining over Nakagami-$m$ fading}},''
  \emph{IEEE Trans. Wireless
  Commun.}, vol.~10, no.~6, pp. 1886--1894, Apr. 2011.
\end{thebibliography}
\end{document}